\documentclass[11pt,floatfix]{revtex4}
\usepackage{psfig,amsmath,amssymb,graphics}
\newcommand {\UC}{U_{\mbox \tiny C}}
\newcommand{\rbar}{\bar{r}}             
\newcommand{\rin}{r_{\rm in}}
\newcommand{\rout}{r_{\rm out}}
\newcommand{\Oin}{\Omega_{\rm in}}
\newcommand{\Oout}{\Omega_{\rm out}}

\newcommand{\etaopt}{\eta_{\rm opt}}
\newcommand{\tmax}{t_{\rm max}}
\newcommand{\Gmax}{G_{\rm max}}
\newcommand{\bu}{\underline{u}}
\newcommand{\bv}{\underline{v}}
\newcommand{\bU}{\underline{U}}
\newcommand{\pd}{\partial}
\newcommand{\Imag}{{\rm Im}}

\begin{document}

\title{Transient growth in Taylor-Couette flow}

\author{Hristina Hristova$^{1,2}$, S\'ebastien Roch$^{1,2}$, Peter
J. Schmid$^{1,3}$, and Laurette S. Tuckerman$^{1,4*}$
\vspace*{.2cm} \\
$^1$Ecole Polytechnique, Palaiseau, France \\
$^2$Ecole Polytechnique de Montr\'eal, Montreal, Canada \\
$^3$University of Washington, Seattle Washington, U.S.A. \\
$^4$LIMSI-CNRS, Orsay, France\\
$^*$Email: laurette@limsi.fr
}

\date{\today}
\maketitle

\section*{Abstract}

Transient growth due to non-normality is investigated for the
Taylor-Couette problem with counter-rotating cylinders as a function of
aspect ratio $\eta$ and Reynolds number $Re.$ 
For all $Re \leq 500$, transient growth is enhanced by curvature,
i.e. is greater for $\eta<1$ than for $\eta=1$, the plane Couette limit.
For fixed $Re < 130$ it is found that the greatest
transient growth is achieved for $\eta$ between the Taylor-Couette
linear stability boundary, if it exists, and one, while for $Re > 130$ 
the greatest transient growth
is achieved for $\eta$ on the linear stability boundary.
Transient growth is shown to be approximately 20\% higher near the 
linear stability boundary at $Re = 310$, $\eta =0.986$ 
than at $Re = 310$, $\eta=1$, near the threshold observed for 
transition in plane Couette flow.
The energy in the optimal inputs is primarily meridional;
that in the optimal outputs is primarily azimuthal.
Pseudospectra are calculated for two contrasting cases.
For large curvature, $\eta=0.5$, the pseudospectra
adhere more closely to the spectrum than in
a narrow gap case, $\eta=0.99$.

\vspace*{2cm}
\noindent
{\bf Keywords:} Taylor-Couette instability, transition to turbulence
\parindent .5cm

\newpage
\section{Introduction}
\label{Introduction}

Two shear-driven flows bear the name of Couette: the flow between
differentially rotating concentric cylinders is called cylindrical
Couette flow, or more commonly, Taylor-Couette flow, while the flow
between infinite parallel plates translating at different velocities
while maintaining a constant separation is called plane Couette flow.
Exact solutions, both called Couette solutions, to the Navier-Stokes
equations for each of these configurations are easily calculated and
serve as illustrations in most textbooks.  Each of these two flows has
become a paradigm of hydrodynamic stability theory.

Taylor-Couette flow can be considered to be a paradigm of understanding.  
In 1923, Taylor \cite{Taylor} carried out calculations of the linear
instability of Couette flow to the onset of axisymmetric vortices and
compared these with experiment, obtaining agreement which remains
remarkable even by today's standards.  In later research, increasingly
ornate and beautiful experimental patterns were discovered,
e.g. \cite{Coles,Andereck,Taggribbons}, 
and correspondingly elaborate numerical,
asymptotic, and theoretical calculations, e.g. 
\cite{Krueger,Marcus,Langford}, reproduced and explained
these patterns, again with remarkable accuracy; 
see \cite{Swinney,DrazinReid,Taggreview,ChossatIooss}.

Plane Couette flow, on the other hand, can be considered to be a
paradigm of mystery.  For plane parallel flows, Squire's theorem
\cite{Squire} firmly establishes that the linear instability with the
lowest critical Reynolds number is spanwise invariant. Armed with this
theorem, researchers have long known, and more recently proved
\cite{Romanov}, that plane Couette flow is linearly stable at all
Reynolds numbers.  Yet, in laboratory experiments \cite{Tillmark} and
in numerical simulations \cite{Lundbladh}, plane Couette flow
undergoes sudden transition to three-dimensional turbulence.  Since at
least the 1960's researchers have explored various mechanisms for
transition in plane Couette flow which bypass linear instability and
Squire's theorem.  An element shared by all of these approaches,
independent of the theoretical mechanism proposed for instability, is
the presence of streamwise vortices 
\cite{Joseph,Butler,Trefethen_93,Reddygrowth,Waleffe}, 
i.e. perturbations which, unlike
those deemed critical by Squire's theorem, are not spanwise invariant,
but are instead invariant or almost invariant in the streamwise
direction.  These are analogous to the Taylor vortices which, in
Taylor-Couette flow, are the eigenvectors responsible for the linear
instability and are realized in experiments and in nonlinear numerical
simulations.

One major line of research has focused on the effect of the non-normality of
the operator governing the linear stability of plane Couette flow
\cite{Butler,Trefethen_93,Reddygrowth}.  Such operators may lead to
transient growth dynamics, even in the absence of linear instability, allowing
nonlinear effects to take over before the final exponential decay exhibited by
the linear evolution.  Plane Couette flow can exhibit transient growth of
several orders of magnitude; the initial conditions which maximize the
transient growth, called optimal perturbations, contain streamwise vortices.
Transient growth is closely associated with the sensitivity of the spectrum to
small perturbations of the operator, quantified by the pseudospectra
\cite{Trefethen_93}.

The concepts of pseudospectra, non-normality, transient growth, and
optimal perturbations, have been rarely applied to
Taylor-Couette flow \cite{Gebhardt,Meseguer}, probably because conventional
linear stability analysis is so successful in explaining the
transitions observed.  However, the large parameter space of
Taylor-Couette flow offers the possibility of tuning the system from a
normal operator, which does not support any transient growth, to a
highly non-normal operator.  Because of translational and Galilean
invariance, plane Couette flow depends only on the distance $2d$
between the plates, the relative velocity $2\Delta U$ of the plates,
and the kinematic viscosity $\nu$, which are combined into a single
nondimensional parameter, the Reynolds number defined conventionally
as $Re \equiv \Delta U d/\nu$.  In contrast, because of curvature
effects in Taylor-Couette flow, both the inner and outer radii $\rin$
and $\rout$ (or gapwidth $2d = \rout-\rin$ and average radius
$\rbar=(\rin+\rout)/2$) play a role.  Because transformation to a
rotating reference frame would introduce a Coriolis term into the
equations, the flow depends on the angular velocities of both the
inner and the outer cylinders, $\Oin$ and $\Oout$.  The five
dimensional parameters can be combined into three non-dimensional
parameters in various ways.  Our choice is $\eta \equiv \rin/\rout$,
$\mu \equiv \Oout/\Oin$, and $Re \equiv \rin\Oin d/\nu$.

The idea of trying to approach the stability of plane Couette flow via
Taylor-Couette flow is an appealing one and has inspired a number of
investigations.  The axisymmetric Taylor-vortex solution undergoes a
secondary bifurcation to a non-axisymmetric wavy-vortex 
solution \cite{Coles,Marcus}.  In
a search for states intermediate in complexity between the Couette
solution and turbulent plane Couette flow, 
Nagata \cite{Nagata90,Nagata98} took
the limits of a narrow gap and almost corotating cylinders and
discovered that, while the Taylor-vortex solution ceases to exist as
the Coriolis term (i.e. the average angular velocity) is decreased,
the wavy-vortex solution could be continued to the plane Couette limit.  
Faisst \& Eckhardt \cite{Eckhardt} showed that these wavy solutions also exist
for counter-rotating cylinders and that the lowest Reynolds number at which
they first appear (via a saddle-node bifurcation) becomes independent of
the rotation ratio as $\eta \rightarrow 1$.
Finally, Prigent and Dauchot \cite{Prigent} have discovered that an analog
of the spiral turbulence state of Taylor-Couette flow also exists
for plane Couette flow.

Our investigation goes somewhat in the opposite direction.  While the
investigations cited above considered steady or traveling states known
to exist in Taylor-Couette flow and continued them to the plane
Couette limit, we take the ideas of pseudospectra, non-normality,
transient growth, and optimal perturbations, apply them to
Taylor-Couette flow and investigate the limit as Taylor-Couette flow
approaches plane Couette flow. We focus on a given representative
azimuthal and axial wavenumber and explore the stability
characteristics of counter-rotating Taylor-Couette flow for a range of
Reynolds numbers and radius ratios. 
In a related study,
Meseguer \cite{Meseguer} studied transient effects -- optimized over
azimuthal and axial wavenumbers -- for a specified radius ratio and
varying Reynolds number and angular velocity ratio.
Since Taylor-Couette flow is linearly stable for $\Oin=0$,
Meseguer suggests that transient growth plays an important role in the
transition to turbulence observed experimentally \cite{Coles}
for $\Oin=0$, $-\Oout \gg 1$,
a theory contested by Gebhardt \& Grossman \cite{Gebhardt}.

Why calculate transient growth in Taylor-Couette flow, 
one of the major success stories of linear stability theory?
Transient growth is a new tool in linear stability theory,
whose worth and longevity have yet to be proved.
It has up to now been applied to flows in which transition
from the basic laminar state is not understood.
It is therefore worthwhile to calculate transient growth
in Taylor-Couette flow, in which transition is believed
to be completely understood by hydrodynamicists.
This could provide information about the value and
significance of transient growth calculations.
The other side of the coin is that the paradigmatic
status of Taylor-Couette flow should be maintained.
As new tools or measurements become available,
they should be brought to bear on this basic flow 
in order for hydrodynamicists to extract additional
knowledge about Taylor-Couette flow.

\section{Governing Equations and Numerical Methods}
\label{Governing Equations and Numerical Methods}

\subsection{Taylor-Couette flow}
\label{Taylor-Couette flow}

The Couette solution is the unique solution of the form $\bU =
\UC(r)\underline{e}_\theta$ to the incompressible Navier-Stokes
equations:
\begin{subequations}
\begin{eqnarray}
\frac{\pd\bU}{\pd t} + (\bU \cdot \nabla) \bU
&=& -\nabla P + \frac{1}{Re}\Delta \bU,\\
\nabla \cdot \bU &=& 0, \\
\bU(r=\rbar-1)=1, 
&\quad& \bU(r=\rbar+1)=\frac{\mu}{\eta}.
\end{eqnarray}
\label{NS}
\end{subequations}
Distances have been nondimensionalized by $(\rout-\rin)/2$ and
velocities by $\rin\Oin$.  We recall that 
\begin{equation}
\eta \equiv \rin/\rout, \quad \mu \equiv \Oout/\Oin, \quad
Re = \rin\Oin (\rout-\rin)/(2\nu) 
\label{redef}\end{equation}
where $\rin$, $\rout$, $\Oin$, $\Oout$ are the inner and outer
cylindrical radii and angular velocities, respectively, and $\nu$ is
the kinematic viscosity.  
Definition (\ref{redef}) of $Re$, chosen for
compatibility with the plane Couette convention for the case $\mu=-1$,
differs by a factor of two from the conventional definition employed
in Taylor-Couette flow.  The average radius is
$\rbar=(1+\eta)/(1-\eta)$.  Thus, the limit $\eta \rightarrow 1$
corresponds to $\rbar \rightarrow \infty$; we refer to either of these
limits as convenient.

The Couette solution is:
\begin{subequations}
\begin{eqnarray}
\UC(r) &=& Ar + \frac{B}{r}, \label{tcstat} \\  
A = \frac{\mu - \eta^2}{2\eta(1+\eta)}, 
&\quad& B = \frac{2\eta (1 - \mu)}{(1-\eta)(1-\eta^2)}. 
\end{eqnarray}
\label{Couette}
\end{subequations}
The expressions for $A$ and $B$ again differ from the standard ones by
factors of two due to our use of $(\rout-\rin)/2$ as unit of length.
Expression (\ref{Couette}) can be viewed as a superposition of solid
body rotation, $Ar$, and the flow due to a point vortex, $B/r$.  As
$\eta \rightarrow 1$, the leading terms of these two contributions
become equal and opposite: $Ar \sim -B/r \sim (\mu-1) \rbar/4$.  For
$\eta \sim 1$, we therefore rewrite (\ref{Couette}) in the following
equivalent form not subject to this cancellation error:
\begin{eqnarray}
\UC(r) = \frac{1}{4 r \rbar(\rbar-1)} 
        \; \left[ 2\rbar^3 \; ((\mu+1)+(\mu-1)y) \right. 
                    + \rbar^2 \; (3(\mu-1)+4(\mu+1)y+(\mu-1)y^2) \nonumber\\
            + \left. 2\rbar y \;((\mu-1)+(\mu+1)y)-(\mu-1)(1-y^2) \right]
\label{nocancel}
\end{eqnarray}
where
\begin{equation}
y\equiv r-\rbar.
\end{equation}

The equations we consider are the Navier-Stokes equations (\ref{NS})
linearized about the Couette solution (\ref{Couette}) or
(\ref{nocancel}):
\begin{subequations}
\begin{eqnarray}
& & \frac{\pd u_r}{\pd t} 
+ \frac{\UC}{r}\frac{\pd u_r}{\pd \theta} 
- \frac{2\UC}{r} u_\theta 
= -\frac{\pd p}{\pd r} 
+ \frac{1}{Re}\Big(\Delta u_r 
- \frac{u_r}{r^2} 
- \frac{2}{r^2}\frac{\pd u_\theta}{\pd \theta}\Big), \label{nscylr}\\
& & \frac{\pd u_\theta}{\pd t} 
+ \frac{\UC}{r}\frac{\pd u_\theta}{\pd \theta} 
+ \frac{\UC}{r} u_r + {\UC}^{'} u_r 
= -\frac{1}{r}\frac{\pd p}{\pd \theta} 
+ \frac{1}{Re}\Big(\Delta u_\theta 
- \frac{u_\theta}{r^2} 
+ \frac{2}{r^2}\frac{\pd u_r}{\pd \theta}\Big), \label{nscylth}\\
& & \frac{\pd u_z}{\pd t} 
+ \frac{\UC}{r}\frac{\pd u_z}{\pd \theta} 
= -\frac{\pd p}{\pd z} 
+ \frac{1}{Re}\Big(\Delta u_z \Big), \label{nscylz}\\
& & \frac{\pd u_r}{\pd r} 
+ \frac{1}{r}u_r 
+ \frac{1}{r}\frac{\pd u_\theta}{\pd \theta} 
+ \frac{\pd u_z}{\pd z} = 0, \label{incomp}
\end{eqnarray}
subject to the boundary conditions 
\begin{equation}
u_r = u_\theta = u_z = 0 \quad {\rm at} \quad r=\rbar\pm 1 \label{cltc}
\end{equation}
\label{navierstokescyl}
\end{subequations}
with
\begin{equation}
\Delta = \frac{\pd^2}{\pd r^2} 
+ \frac{1}{r}\frac{\pd}{\pd r} 
+ \frac{1}{r^2}\frac{\pd^2}{\pd \theta^2} 
+ \frac{\pd^2}{\pd z^2}. \nonumber 
\end{equation}
\addtocounter{equation}{-1}

The Taylor-Couette geometry and the Couette solution are homogeneous in the
azimuthal ($\theta$) and axial ($z$) direction, which are analogous to the
streamwise ($x$) and spanwise ($z$) directions in plane Couette flow.
Table~\ref{tableEquiv} summarizes the equivalence of the coordinate systems
for plane Couette and Taylor-Couette flow. Solutions to
(\ref{navierstokescyl}) which are spatially bounded are therefore
trigonometric in each of these directions, with wavenumbers $m$ and $\beta,$
\begin{equation}
(u_r, u_\theta, u_z, p) = (\hat{u}_r, \hat{u}_\theta, 
\hat{u}_z, \hat{p})(r,t) \exp (im\theta + i\beta z). 
\end{equation}

The parameter space is too vast to permit full exploration.  In this
study, we limit ourselves to $\mu=-1$, $m=0$, and $\beta =\pi/2$.  We
choose $\mu=-1$ so that the average angular velocity vanishes.

The choice $m=0$ of axisymmetric perturbations is made for simplicity.
In plane Couette flow, the transient growth achieved by
streamwise-independent perturbations is, while not maximal, very close
to the optimal value.  Although the study of transient growth in 
Taylor-Couette flow is relatively unexplored, its linear instability has
been extensively studied, e.g.
\cite{Taylor,Krueger,Langford,DrazinReid,Gebhardt}.  
The linear instability undergone by
Taylor-Couette flow with counter-rotating cylinders is usually
non-axisymmetric and leads to spirals rather than vortices.  However,
the threshold associated with axisymmetric perturbations is very close
to the actual non-axisymmetric threshold.  Indeed, the thresholds are
so close that the instability was thought to be axisymmetric until
1966, when calculations by Krueger, Gross and DiPrima \cite{Krueger}
confirmed experimentally by Coles \cite{Coles}
showed the first instability to be non-axisymmetric for $\mu$
sufficiently negative, more precisely $\mu \lesssim 0.78$ in the
narrow-gap limit.
Note that the correspondence between the streamwise wavenumber $\alpha$ 
of plane
Couette flow and the azimuthal wavenumber $m$ of Taylor-Couette flow
is $m \sim \rbar \alpha$, since $x \sim r\theta$.  Setting $m$ to
correspond to a fixed non-zero value of $\alpha$ would thus require
increasing $m$ through integer values as $\eta \rightarrow 1$ or
$\rbar \rightarrow \infty$.

The choice $\beta=\pi/2$ corresponds to an axial wavelength of 4.
Since our choice of length scales dictates that the radial gap is of
width 2, this means that a single vortex has the same axial as radial
extent, i.e. is approximately circular.  The axial wavelength
corresponding to linear instability is somewhat smaller than this for
counter-rotating cylinders by as much as 20\%, but is nonetheless close.

We eliminate $\hat{u}_z$ and $\hat{p}$ by using (\ref{nscylz}) and the
condition of incompressibility (\ref{incomp}), obtaining evolution
equations in $\bu = (\hat{u}_r,\hat{u}_\theta)$.  The perturbations
$(\hat{u}_r,\hat{u}_\theta)$ are represented as series of Chebyshev
polynomials in $y \equiv r-\rbar$, which are evaluated at the
Gauss-Lobatto points \cite{Canuto}.

The evolution equation is written symbolically for the state vector
$\bu = (\hat{u}_r,\hat{u}_\theta)^T$ as
\begin{equation}
\frac{\pd}{\pd t} \bu 
= -i {\cal \hat{L}} \bu.
\label{evolnumer}
\end{equation}
The axial velocity component $\hat{u}_z$ is calculated from
$(\hat{u}_r,\hat{u}_\theta)$ via incompressibility.


\begin{table}
\begin{center} 
\begin{tabular}{|c|c|c||c|c|c|}
\hline
\multicolumn{3}{|c||} {Plane Couette} & \multicolumn{3}{|c|} {Taylor-Couette} \\
\hline \hline
$x$ & streamwise & $\alpha$ & $\theta$ & azimuthal & $m/\rbar$\\ \hline
$z$ & spanwise & $\beta$ & $z$ & axial & $\beta$ \\ \hline
$y$ & normal & & $r$ & radial & \\ \hline
\end{tabular} \label{tab:criticReynolds}
\caption{Equivalence of coordinate systems for plane Couette and
  Taylor-Couette flow.}
\end{center}
\label{tableEquiv}
\end{table}

Our codes for discretizing the Taylor-Couette operator were based on a
pseudospectral representation of the radial and azimuthal velocity in the
inhomogeneous normal direction, similar to the Matlab code for plane Couette
flow written by Reddy \cite{Reddygrowth,Reddypseudo} and published in
\cite{Schmid}.  We tested our Taylor-Couette code in a number of ways.  First,
we took the limit $\eta \rightarrow 1$ while maintaining a fixed Reynolds
number and verified that we obtained the same asymptotic growth rates as in
plane Couette flow.  We then took $\eta \rightarrow 1$ and verified that the
critical Reynolds numbers were those obtained by Krueger et al.
\cite{Krueger}.  Finally, for arbitrary values of $\eta$, we verified that the
critical Reynolds numbers were those given in \cite{Swinney}.

\subsection{Optimal Growth and Pseudospectra}

As a measure of growth, we use the energy norm defined by:
\begin{equation}
E(\bu) = \|\bu\|^2_E = \underbrace{
\int_{r=\rbar-1}^{\rbar+1} \left( \vert
\hat{u}_r \vert^2 + \frac{1}{\beta^2}
\vert {\cal{D}} \hat{u}_r \vert^2 + \frac{m^2}{\beta^2 r^2}\vert
\hat{u}_\theta \vert^2 \right)\; r \; dr}_{E_{r,z}} +  
\underbrace{
\int_{r=\rbar-1}^{\rbar+1} \vert \hat{u}_\theta \vert^2 \; r \; 
dr}_{E_\theta}. 
\label{energy}
\end{equation}
with ${\cal{D}}f \equiv \frac{1}{r}\frac{d}{dr}(rf).$ 
The energy norm is particularly significant in hydrodynamics,
because the nonlinear terms of the Navier-Stokes equations conserve energy. 
This lends significance to the study of the linearized equations 
(\ref{navierstokescyl}), since any energy growth that takes place must occur 
via linear mechanisms \cite{Schmid}. 
The maximal energy growth at time $t$ for $\bu$ evolving according 
to (\ref{evolnumer}) is defined by:
\begin{equation}
G(t) \equiv \sup_{\bu(0)\neq 0} 
\frac{\|\bu(t)\|^2_E}{\|\bu(0)\|^2_E}
=\|\exp(-i{\hat{\cal{L}}}t)\|^2_E.
\label{Gt}\end{equation}
Thus the maximal energy growth is given by the energy norm of the
operator $\exp(-i{\hat{\cal L}}t)$.

The norm of a normal operator is its dominant eigenvalue $\lambda_{\rm
max}$; this is the largest factor by which matrix multiplication can
increase the norm of a vector.  For a non-normal operator, cross-terms
between non-orthogonal eigenvectors typically contribute to the norm
of a vector.  Instead, it is the singular vectors which are
orthogonal; the norm of the operator is given by the largest singular
value $\sigma_{\rm max}$.  The singular value $\sigma_{\rm max}$ and
its corresponding normalized right and left singular vectors $\bu_{\rm
max}$, $\bv_{\rm max}$ satisfy:
\begin{equation}
\exp(-i{\hat{\cal L}}t) \bu_{\rm max} = \sigma_{\rm max} \bv_{\rm max}
\label{singval}\end{equation}
i.e. linear evolution from initial condition $\bu_{\rm max}$
leads to the state $\sigma_{\rm max} \bv_{\rm max}$.

The statements above are all inner-product dependent: an operator is
normal or not and vectors are orthogonal or not with respect to a
particular inner product.  The singular value decomposition is
inner-product dependent as well.  Since we investigate growth in the
energy norm, we seek the largest singular value and its corresponding
left and right singular vectors in the energy norm as well.  Standard
software, however, provides singular value decompositions with respect
to the 2-norm.  Additionally, each of the values on the Gauss-Lobatto
grid points must be multiplied by a weight appropriate for calculating
the energy (\ref{energy}). We compute the elements $M_{ij}$ of an $N \times N$
Hermitian matrix ${\bf{M}}$ by taking the inner products, derived from
(\ref{energy}), between two eigenfunctions $\Phi_i$ and $\Phi_j$ of
$\hat{\cal{L}}.$ The Cholesky factorization of ${\bf{M}} = {\bf{F}}^H
{\bf{F}}$ is then used to convert the energy norm of the operator exponential
to an $L_2$-norm of the weighted matrix exponential according to 

\begin{equation}
\|\exp(-i{\hat{\cal L}} t)\|^2_E \approx \| {\bf{F}} \exp(-i{\bf{D}} 
t) {\bf{F}}^{-1}\|^2
\label{weight}
\end{equation}
with ${\bf{D}}$ as an $N \times N$ diagonal matrix consisting of the
eigenvalues of $\hat{\cal{L}}$ (see \cite{Schmid} for more details). We
calculate the largest singular value (under the 2-norm) of the operator on the
right-hand-side of (\ref{weight}). Its square is the maximal energy growth.
The number $N$ of eigenfunctions has been chosen large enough to ensure
converged results.

The optimal growth is defined as
\begin{equation}
\Gmax = \sup_{t\geq 0} G(t).
\label{Gmax}\end{equation}
If $\hat{\cal L}$ has an eigenvalue with positive imaginary part, then
$\Vert \bu \Vert$ grows exponentially in time (for any norm) and so
$\Gmax=\infty$.  Thus calculations of optimal growth are meaningful
only for operators which are linearly stable.

We also wish to keep track of the quantities responsible for achieving
the maxima in (\ref{Gt}) and (\ref{Gmax}).  The time for optimal
growth $\tmax$ is that which achieves the maximum (sup) in
(\ref{Gmax}), i.e. $\Gmax=G(\tmax)$.  The optimal input, denoted by
$\bu(0)$, is the normalized initial condition which achieves the
maximum (sup) in (\ref{Gt}) for $t=\tmax$.  The optimal output,
denoted by $\bu(\tmax)$, is the velocity field resulting from the
linearized Taylor-Couette evolution (\ref{evolnumer}) starting from
the unit-energy optimal input $\bu(0)$; its energy gain is $\Gmax$.

A non-normal operator ${\cal L}$ is also characterized by its
pseudospectra \cite{Trefethen_93}.  The $\epsilon$-pseudospectrum
$\Lambda_\epsilon({\cal L})$ is the set of complex values $z$
(parametrized by $\epsilon$) which satisfies the property:

\begin{equation}
\|(zI-{\cal L})^{-1}\|\geq\epsilon^{-1}
\label{defpseudo}\end{equation}
Equivalent definitions of the $\epsilon$-pseudo\-spec\-trum are given
in \cite{Trefethen_93,Reddygrowth,Reddypseudo,Schmid,Trefethen_99}.
The definition of the pseudospectra, like that of transient growth,
depends on the norm or inner product.  For
a normal operator, the $\epsilon$-pseudospectrum is the union of the
balls of radius $\epsilon$ surrounding each eigenvalue. For a
non-normal operator, on the other hand, the $\epsilon$-pseudospectrum
may be much larger.

{\it Kreiss' theorem} relates the optimal growth and the pseudospectra
by the following inequality:
\begin{equation}
\Gmax\geq
\sup_{\epsilon>0}\; \left(\epsilon^{-1}
\sup_{z \in \Lambda_\epsilon({\cal L})} \Imag(z)\right).
\label{Kreiss}\end{equation}
The right-hand-side of (\ref{Kreiss}) maximizes (over all strictly
positive values of $\epsilon$) the ratio of the distance to the real
axis of any point in the $\epsilon$-pseudospectrum $\Lambda_\epsilon$
to the value of $\epsilon.$ In practice, it is found, both for
Taylor-Couette flow and for plane channel flows, that the optimal
growth given by the left-hand-side of (\ref{Kreiss}) is approximately
twice the lower bound given by the right-hand-side of (\ref{Kreiss}).

For computing the optimal growth, we used the Matlab code written by
Reddy \cite{Reddygrowth,Reddypseudo}, given in \cite{Schmid}.  For
computing the pseudospectra, we used the code {\tt EigTool}
written by Wright \cite{Wright}, which in turn makes use of the algorithm
developed by Trefethen \cite{Trefethen_99} and is available at website 
{\tt http://www.comlab.ox.ac.uk/projects/pseudospectra}.

\section{Results}

\begin{figure}
\psfig{file=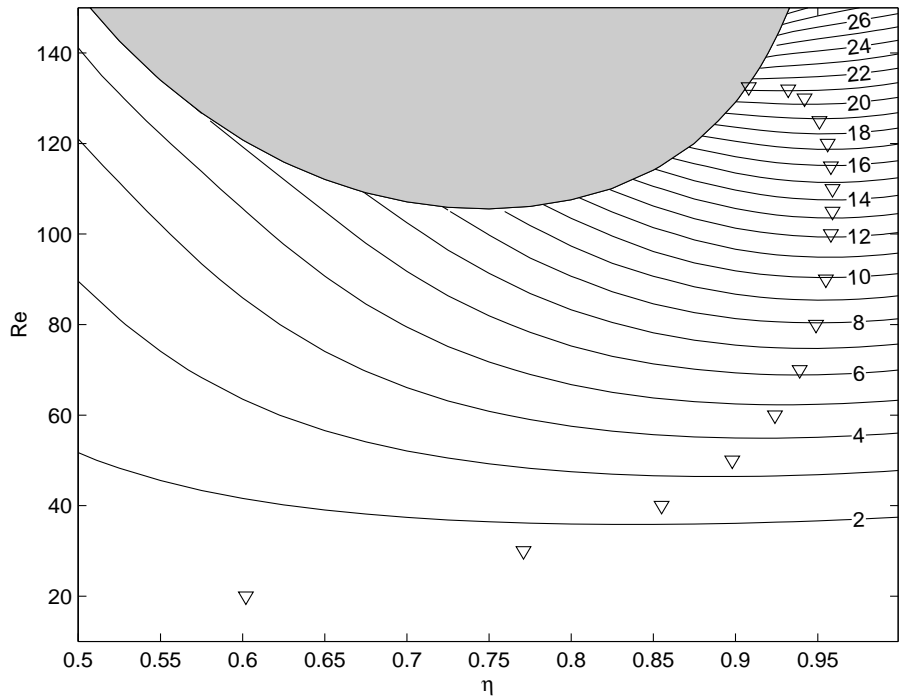,width=12cm}
\caption{Contours of optimal growth for Taylor-Couette flow
in the $(\eta,Re)$ plane.
Shaded area indicates the region of linear instability.
Triangles indicate $\eta_{\rm opt}(Re)$, the value of $\eta$
at which maximum growth is attained for a given value of Re.}
\label{tcoptgr}
\end{figure}

\begin{figure}
\psfig{file=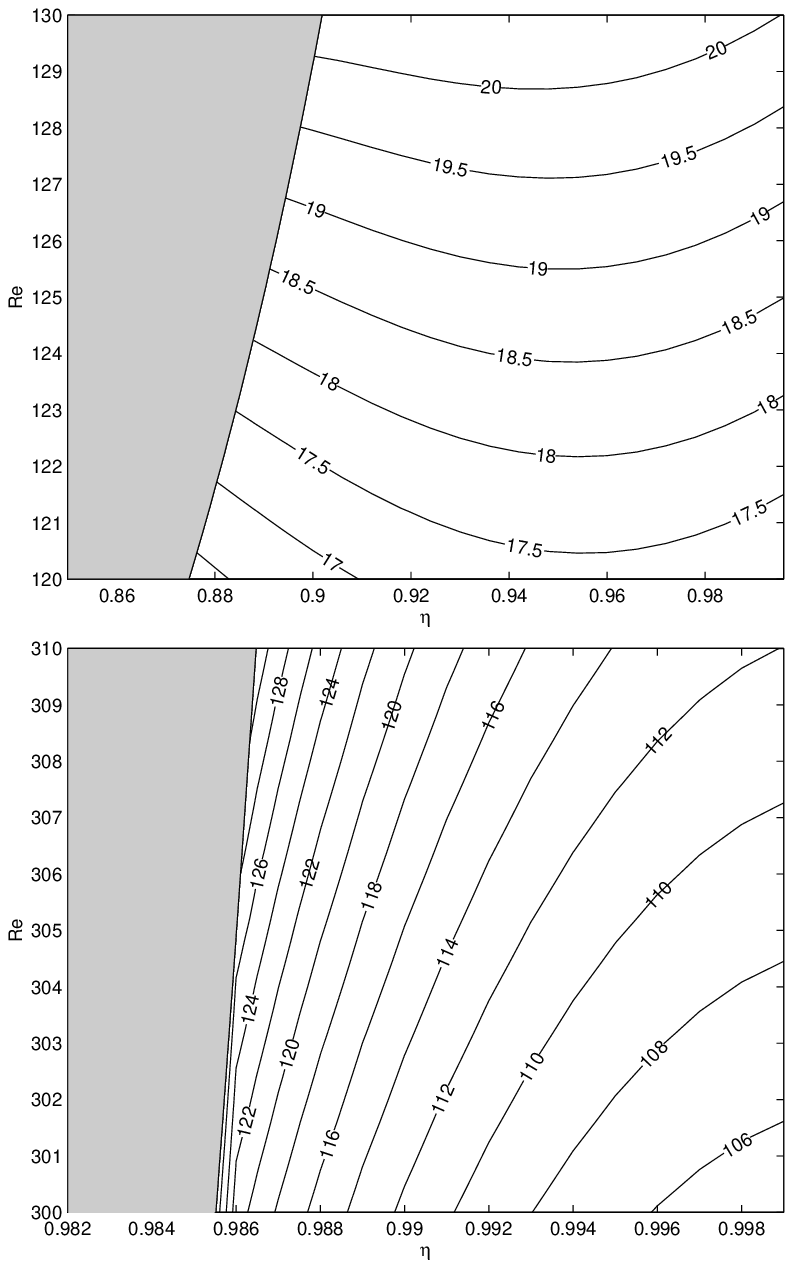,width=10cm}
\caption{Close-up views of contours of optimal growth.
Above: for $120 \leq Re \leq 130$, each contour changes slope at 
a value $\etaopt(Re) \approx 0.96$.
Below: for $Re > 130$, $\Gmax$ increases as $\eta$
decreases to the linear stability boundary $\eta_L$.}
\label{closeup}
\end{figure}

We begin by presenting the optimal growth as a function of $\eta$ and
$Re$.  We recall that throughout our study, we fix the azimuthal
wavenumber $m=\alpha=0$, the axial wavenumber $\beta=\pi/2$, and the
angular velocity ratio $\mu=-1$.  Figure \ref{tcoptgr} shows the
contours of constant optimal growth $\Gmax$.  Inside the shaded
region, Couette flow is linearly unstable to axisymmetric
perturbations.  The boundary of this region is the critical Reynolds
number $Re_L(\eta)$.  $Re_L \rightarrow \infty$ as $\eta \rightarrow
1$, as expected since plane Couette flow is linearly stable for all
Reynolds numbers.  We may also consider the rightmost portion of the
linear stability boundary as a function $\eta_L(Re)$.

A striking feature is that the maximum growth for a fixed Reynolds
number, indicated by the triangles in figure \ref{tcoptgr}, is always
achieved for a radius ratio $\eta=\etaopt(Re)$ which is less than one.
We propose a possible explanation for this trend.
As we decrease the radius ratio $\eta$, the asymptotic growth rate,
i.e. the imaginary part of the least stable eigenvalue, increases. At
the same time, the non-normality of the operator decreases, resulting
in diminished transient growth. The combination of these effects
results in a maximum growth rate that is achieved for values of $\eta$
less than one.  We see that $\etaopt \approx 0.6$ for $Re=20$,
increases to a maximum of $\etaopt \approx 0.96$ for $Re \approx 110$,
and then abruptly decreases and terminates by meeting the linear
instability boundary $\eta_L \approx 0.9$ at $Re \approx 130$.  For
$Re > 130$, the maximum growth is achieved for $\eta=\eta_L(Re)$.  
The enlargements in figure \ref{closeup} show the
typical behavior of the contours of $\Gmax$ for $Re$ near 130 and for
$Re$ near 300.  For $Re \leq 130$, the optimal growth is fairly weak,
varying from a factor of 1 to 21.  Arbitrarily high values of $\Gmax$
can be attained by increasing $Re$, since for plane Couette flow,
i.e. $\eta=1$, it is known \cite{Reddygrowth} that $\Gmax \sim Re^2$.
In fact, over the range $300 < Re < 310$, $\Gmax$
is approximately 20\% higher for $\eta=\eta_L$ than for $\eta=1$.  
In this range, the $\Gmax$ contours are nearly vertical as they 
approach the linear instability boundary, meaning that $\Gmax$ is far more
sensitive to a decrease in $\eta$ than to an increase in $Re$.  
It is near $Re=310$ that plane Couette flow undergoes a sudden 
unexplained transition to turbulence.
Table \ref{tabconv} gives selected numerical values of $\Gmax$.

\begin{table}
\begin{center}
{\small
\texttt{
\begin{tabular}{|l||c|c|c|c|c|c|c|c|c||c|}
\hline
\hspace{1cm} $\eta$ & 0.5 & 0.6 & 0.7 & 0.8 & 0.9 & 0.95 & 0.99 & 0.999 & 0.9999 & Plane\\
$Re$ & & & & & & & & & & Couette \\
\hline
\hline
50 & 1.95 & 2.39 & 2.85 & 3.23 & 3.38 & 3.34 & 3.26 & 3.23 & 3.23 & 3.23 \\
\hline
75 & 2.62 & 3.51 & 4.66 & 5.95 & 6.90 & 7.02 & 6.90 & 6.85 & 6.84 & 6.84 \\
\hline
100 & 3.29 & 4.69 & 6.74 & 9.40 & 11.70 & 12.14 & 12.01 & 11.93 & 11.92 & 11.92 \\
\hline
125 & 4.17 & --- & --- & --- & 18.45 & 18.87 & 18.59 & 18.46 & 18.44 & 18.44 \\
\hline
150 & 5.53 & --- & --- & --- & --- & 27.72 & 26.65 & 26.44 & 26.42 & 26.42 \\
\hline
300 & --- & --- & --- & --- & --- & --- & 111.60 & 104.87 & 104.74 & 104.73 \\
\hline
500 & --- & --- & --- & --- & --- & --- & --- & 291.92 & 290.39 & 290.35 \\
\hline
\end{tabular}
}}
\caption{Optimal growth $\Gmax$ in the narrow-gap limit.
Dashed entries indicate linear instability, i.e. $\Gmax=\infty$.
Note that, for fixed $Re$, the maximum optimal growth is achieved
for $\eta < 1$.}
\label{tabconv}
\end{center}
\end{table}

\begin{figure}
\includegraphics{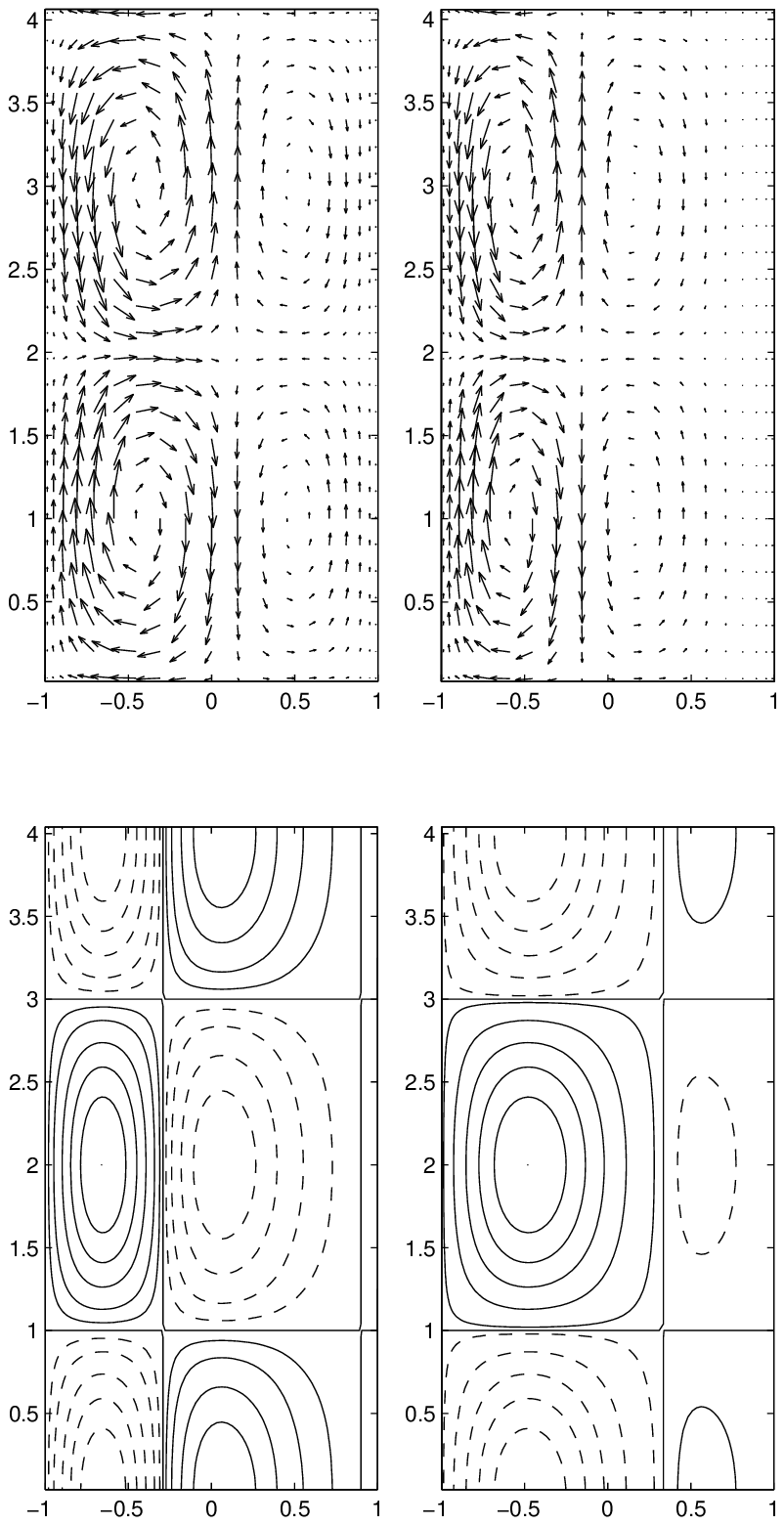}
\caption{Optimal perturbations for $\eta=0.50$, $Re=125$. 
Left: optimal input. Right: optimal output.
Above: Meridional velocity field $(u_r,u_z)$. Below: contours of $u_\theta$.
The energy of the input (output) is primarily in the meridional 
(azimuthal) components. For this reason, arrow lengths and
contour levels are scaled differently for the input and the output.
}
\label{tcpert50}
\end{figure}
 
\begin{figure}
\includegraphics{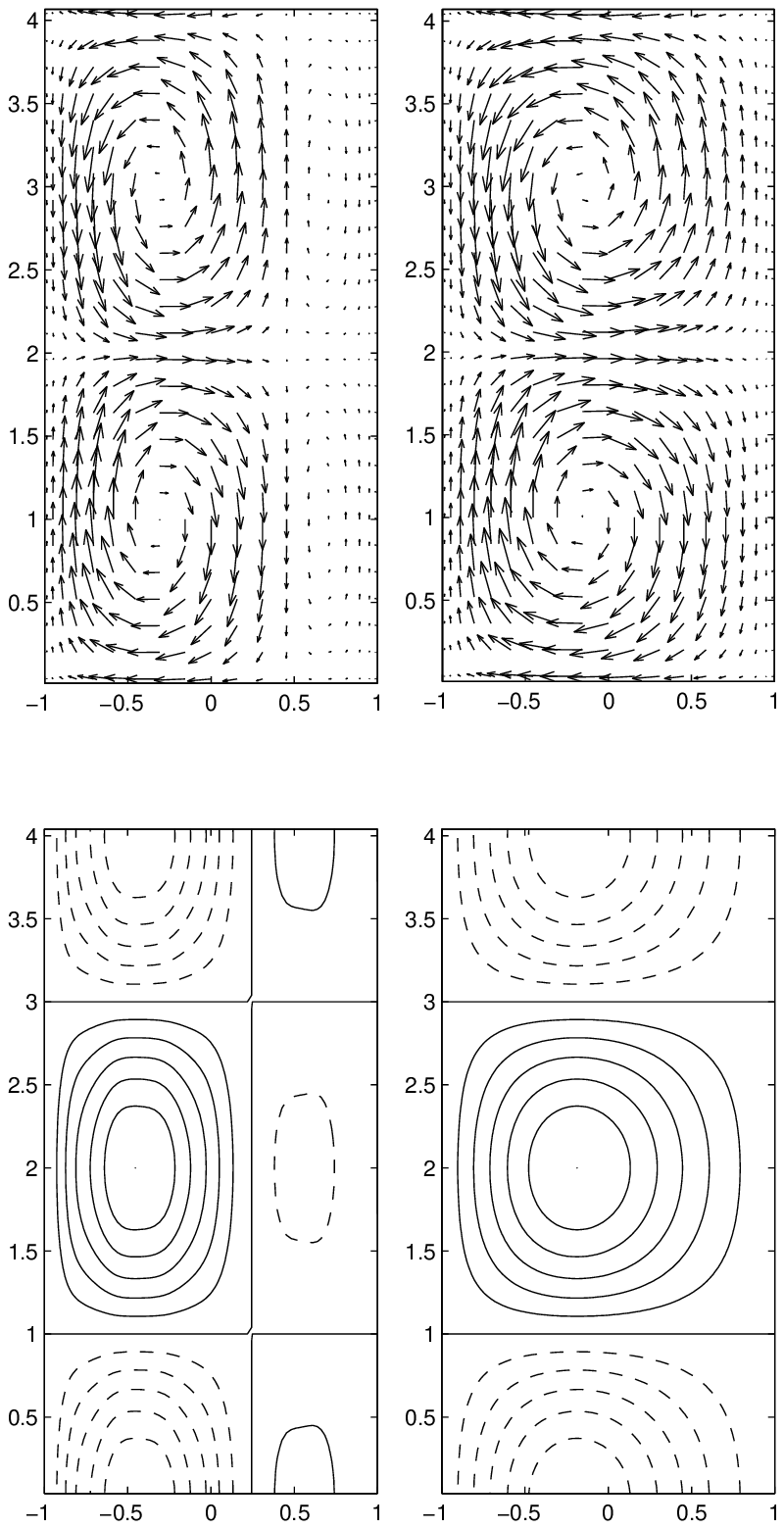}
\caption{Optimal perturbations for $\eta=0.99$, $Re=350$. 
Left: optimal input. Right: optimal output.
Above: Meridional velocity field $(u_r,u_z)$. Below: contours of $u_\theta$.
The energy of the input (output) is primarily in the meridional 
(azimuthal) components; arrow lengths and
contour levels are scaled differently for the input and the output.
}
\label{tcpert99}
\end{figure}

\begin{figure}
\includegraphics{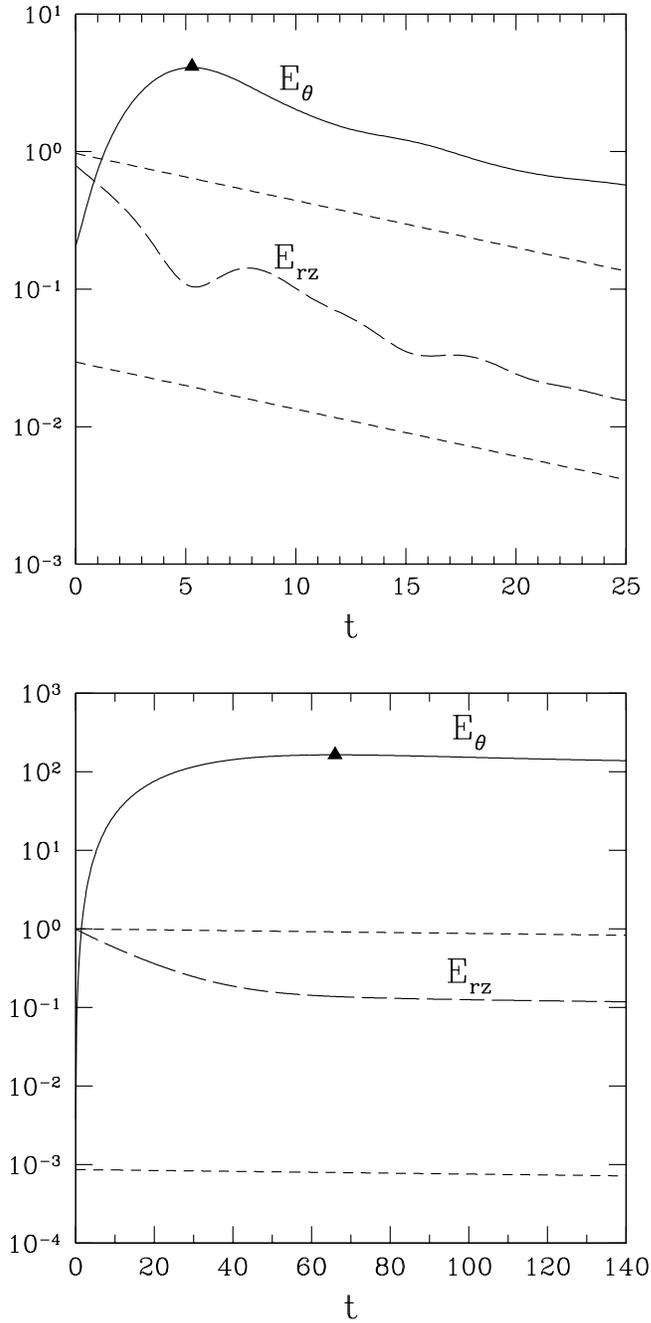}
\caption{Time evolution of optimal growth.
Solid curves denote azimuthal energy $E_\theta(t)$, long-dashed curves
meridional energy $E_{r,z}(t)$ during evolution
from optimal input $\bu(0)$.
Higher and lower short-dashed curves represent
$E_\theta(t)$ and $E_{r,z}(t)$, respectively, for
least stable eigenvector. Triangle corresponds to $\tmax$.
Above: $\eta=0.50$, $Re=125$. Below: $\eta=0.99$, $Re=350$.}
\label{timeseries}
\end{figure}

\begin{figure}
\includegraphics{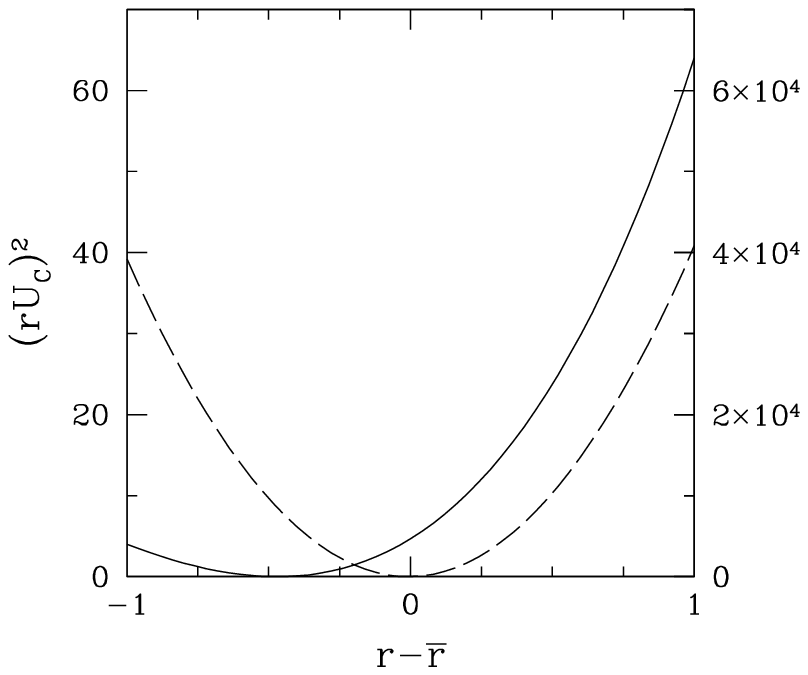}
\caption{Square of angular momentum of basic Couette solution $(r\UC)^2$.
Solid curve: $\eta=0.5$, axis on left.
Dashed curve: $\eta=0.99$, axis on right. 
Rayleigh's criterion for inviscid instability states that 
$\UC$ is unstable where $(r\UC)^2$ is a decreasing function of $r$.}
\label{tcangmom}
\end{figure}

We now study in detail two contrasting cases:
$\eta=0.5$, $Re=125$ and $\eta=0.99$, $Re=350$. 
In the first case, $\eta=0.5$ or equivalently $\rbar=3$,
curvature obviously plays an important role.
The second case, $\eta=0.99$ or equivalently $\rbar=199$,
is very near the plane Couette limit.
The Reynolds numbers have been chosen to be close to the
linear instability threshold $Re_L$ in each case 
in order to maximize transient growth while remaining
within the linearly stable region.

Figures \ref{tcpert50} and \ref{tcpert99} show
the optimal input $\bu(0)$ and output $\bu(\tmax)$ for each case.
The least stable eigenvector, i.e. that with the smallest
decay rate, is not shown, but resembles the optimal output.
The upper portion of each figure shows the meridional velocity
fields $(u_r,u_z)$ of the optimal input and output, 
while the lower portion shows contours of azimuthal velocity $u_\theta$.
The meridional velocity fields consist of vortices whose axes are 
oriented in the azimuthal direction, similar to the eigenvectors which
lead to Taylor vortices at slightly higher Reynolds numbers
and to the streamwise vortices which are the optimal
inputs in plane Couette flow.
The azimuthal components of the optimal inputs and outputs,
shown in the lower portions of figures \ref{tcpert50} and \ref{tcpert99},
are in phase opposition with the vortices, 
with nodal lines at $z=1,3$ going through the vortex centers.

Figure \ref{timeseries} shows the evolution in time of the energies in
the meridional components $E_{r,z}$ and in the azimuthal component 
$E_\theta$ starting from the optimal input $\bu(0)$.  
While the optimal inputs $\bu(0)$ are concentrated primarily in the
meridional components, it is the azimuthal component which dominates
the optimal outputs $\bu(\tmax)$.
For this reason, in order to show
the qualitative geometric features of the two fields,
the inputs and outputs of Figures \ref{tcpert50} and \ref{tcpert99} 
use different scales for the arrow lengths and for the contour levels. 
This evolution corresponds
to the generation of streaks -- deformations of the azimuthal velocity
profile -- by the vortices.  This physical process, referred to as the
lift-up mechanism, has been described, e.g., in \cite{liftup} and is
believed to be a key element in the transition to turbulence in plane
Couette flow.

The graph on the left of figure \ref{timeseries}
shows the evolution of $E_\theta$ and $E_{r,z}$ 
starting from the optimal input $\bu(0)$
and from the least stable eigenvector for $\eta=0.50$, $Re=125$.
The initial energies are $E_\theta(0)=0.2$ and
$E_{r,z}(0)=0.8$.  Initially, over $0 \leq t \leq \tmax \approx 5$,
$E_\theta$ rises while $E_{r,z}$ decreases, attaining values of
$E_\theta(\tmax) \approx 4$ and $E_{r,z}(\tmax) \approx 0.1$, with a
ratio $E_\theta/E_{r,z}(\tmax) \approx 40$.  Over the interval $5 \leq
t \leq 8$, $E_{r,z}$ increases, while $E_\theta$ continues to
decrease.
With further evolution, both energies decrease as $\bu(t)$ converges
towards the least stable eigenvector. From their values at $t=20$, we
estimate $E_\theta/E_{r,z}(\infty) \approx 0.73/0.023 \approx 32$.

The energy evolution for the case $\eta=0.99$, $Re=350$, shown on
the right of figure \ref{timeseries}, resembles that
for plane Couette flow.  The optimal input is almost exclusively
meridional, with negligible azimuthal component: $E_{r,z} \approx
0.996$ while $E_\theta(0) \approx 0.004.$ By $\tmax \approx 66$,
$E_\theta$ has increased to 164 and $E_{r,z}$ decreased to 0.14, a
ratio $E_\theta/E_{r,z}(\tmax)$ of 1170 for the optimal output. This
ratio is approximately maintained as both energies slowly decrease
during the evolution of $\bu(t)$ towards the least stable eigenvector.

In the case $\eta=0.5$, $Re = 125$, two arrays of
vortices are present in the optimal input,
a larger and stronger array near the inner cylinder and 
a smaller and weaker array near the outer cylinder.
Three arrays are present in the optimal output,
whose radial extent and strength decreases in
going from the inner to the outer cylinder.
In the case $\eta=0.99$, $Re = 350$, the optimal
input contains one array of vortices and the optimal
output contains a second weaker, narrower array near
the outer cylinder.
Some light can be shed on the form of these perturbation
fields and on the difference between the two cases
by Rayleigh's criterion for instability in Taylor-Couette flow.

Rayleigh's argument, valid for inviscid and axisymmetric flow, is that
perturbations interchanging rings of fluid at different radii
(e.g. Taylor vortices) will be favored or opposed by the ambient
pressure gradient, according to whether the square of the angular
momentum $(r\UC)^2$ decreases or increases radially outwards.  For
counter-rotating cylinders ($\mu<0$), the sign of $d(r\UC)^2/dr$
changes within the gap.  Rayleigh's criterion is then applied to argue
that only the inner portion of the gap is unstable. This modification
is justified by three related tendencies \cite{DrazinReid}.  First,
the unstable eigenvector is concentrated near the inner cylinder,
where $d(r\UC)^2/dr$ is negative.  Second, the axial wavelength
corresponding to the most unstable or least stable eigenvector
decreases, favoring vortices which remain closer to circular.  Third,
the critical Reynolds number for linear instability increases, meaning
that the critical Reynolds number based on the unstable portion of the
gap remains nearly constant.

Figure \ref{tcangmom} shows that the square of the angular 
momentum decreases radially outwards over the interval 
$r \lesssim \rbar -0.5$ for $\eta=0.50$
and over the interval $r \lesssim \rbar$ for $\eta=0.99.$ 
Although exact application of Rayleigh's criterion would lead to
optimal perturbations far more concentrated near the inner cylinder
than they actually are, the criterion provides a heuristic explanation
for the asymmetry.  Rayleigh's criterion is usually invoked to explain
linear instability, i.e. exponential growth.  
However, a modified version of the criterion should apply
to transient growth as well. 

\begin{figure}
\includegraphics{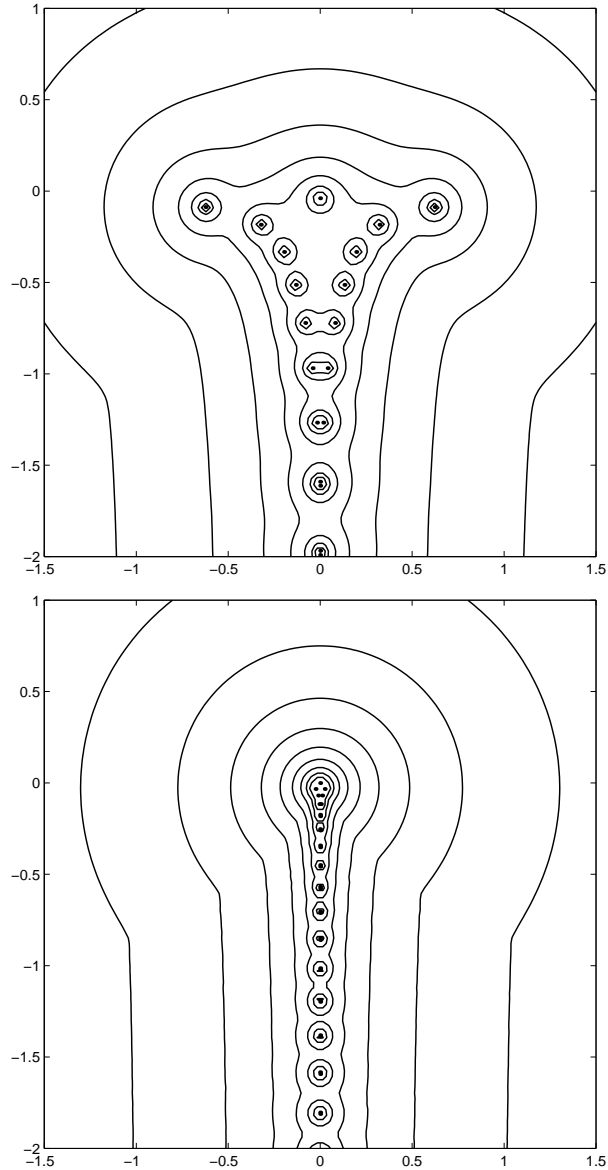}
\caption{Pseudospectra for Taylor-Couette flow.
Above: $\eta = 0.5$ and $Re = 125$.
Contours correspond to $\epsilon=10^{-1.8}, 10^{-1.5},\ldots, 10^{0}$.
Below: $\eta = 0.99$ and $Re = 350$.
Contours correspond to $\epsilon=10^{-2.4}, 10^{-2.1},\ldots, 10^{0}$.}
\label{tcpseudo}
\end{figure}

Finally, we show the spectrum and pseudospectra of the operators
${\cal L}$ for the two cases in figure \ref{tcpseudo}.  
Both pseudospectra plots contain
contours (for the smaller values of $\epsilon$) surrounding individual
eigenvalues and bulb-shaped contours (for the larger values of
$\epsilon$) surrounding the entire spectrum.  For both values of
$\eta$, the contours for a fixed small value of $\epsilon$ surrounding
the eigenvalues near the real axis are wider than the contours
surrounding the eigenvalues farther from the real axis. The
bulb-shaped contours, however, differentiate between the two values of
$\eta$.  The spectrum for $\eta=0.5$ contains eigenvalues near the
real axis with real parts extending to approximately $\pm 0.7;$ a
bulb-shaped pseudospectral contour thus remains a fairly constant
distance from the spectrum.  The spectrum for $\eta=0.99$ is, in
contrast, quite localized on the imaginary axis; in this case, a
bulb-shaped pseudospectral contour protruding into the unstable
half-plane is a first indication of non-normal effects.

We use more detailed calculations of the pseudospectra to compute
approximations to the lower bound (\ref{Kreiss}) on optimal growth of
Kreiss' theorem.  For $\eta=0.5$, we obtain an upper bound
$\Imag(z)\leq 0.1$ for the $\epsilon$-pseudospectrum with $\epsilon =
10^{-1.15}$, which yields the lower bound $\Gmax \geq (10^{1.15}
\times 0.1)^2 = 1.99$, about half of the exact value $\Gmax = 4.17$
that we have calculated.  For $\eta=0.99$, we obtain an upper bound
$\Imag(z)\leq 0.03$ for the $\epsilon$-pseudospectrum with $\epsilon =
10^{-2.45}$, which yields the lower bound $\Gmax \geq (10^{2.45}
\times 0.03 )^2 = 71.49$, again about half of the exact value of 155.

\section{Conclusions}

We have calculated the pseudospectra and optimal transient growth
for Taylor-Couette flow.

Our major result is that, for a fixed Reynolds number, the
optimal transient growth is achieved for a radius ratio $\eta =
\etaopt < 1$, rather than for the plane Couette limit $\eta = 1$. This
is due to the combined effect of increasing modal growth and
decreasing non-normality. As shown in figure 
\ref{tcoptgr}, for $Re<130$, the optimal transient growth
for fixed $Re$ occurs at an intermediate value of $\eta$ ranging from
$\etaopt=0.6$ at $Re=20$ to $\etaopt=0.96$ at $Re=110$.  For $Re >
130$, the optimal transient growth for fixed $Re$ increases as $\eta$
is decreased from the plane Couette limit $\eta = 1$ to the linear
stability boundary $\eta=\eta_L$, as shown in figure \ref{closeup}.

If transient growth on the order of $\Gmax\approx 100$ is 
suspected to initiate transition in plane Couette flow near $Re = 300$,
then transition should also occur near $Re = 300$
outside the linear instability domain in Taylor-Couette flow, 
albeit in very narrow gaps ($\eta \gtrsim 0.986$).
If these predictions fail to hold, then theories concerning
transition initiated by transient growth must be modified
or even abandoned.
Another possible direction for future studies concerns 
regions of $(\eta, Re)$ in which transient growth
competes with linear instability;
such situations should be included in a comprehensive
theory of transition initiated by non-normal effects.

Two cases have been considered in detail: 
$\eta=0.99$, $Re=350$ and $\eta=0.5$, $Re=125$.
Both show transient growth.
As could be expected, the case $\eta=0.99$, $Re=350$,
shows much higher transient growth, both because of its
greater resemblance to plane Couette flow and also because
of its higher Reynolds number.
As shown in figure \ref{timeseries},
for $\eta=0.99$, $Re=350$, energy grows from the
optimal input by a factor of $\Gmax=164$,
while for $\eta=0.5$, $Re=125$, we have $\Gmax=4$.

As is the case for plane Couette flow, the physical mechanism
accompanying transient growth consists of the conversion
of vortices (here, termed azimuthal rather than streamwise)
into streaks, i.e. perturbations of the basic Couette profile.
This is seen in the ratio of azimuthal to meridional energy
in figure \ref{timeseries}
of the optimal inputs $\bu(0)$ and outputs $\bu(\tmax)$.
The optimal outputs,
depicted in figures \ref{tcpert50} and \ref{tcpert99},
are of higher amplitude near the inner
cylinder, especially in the case $\eta=0.5$, $Re=125$,
as indicated by Rayleigh's criterion for centrifugal instability
illustrated in figure \ref{tcangmom}.

Although both sets of pseudospectra in
figure \ref{tcpseudo} show signs of non-normality,
those for $\eta=0.99$, $Re=350$ show more deviation from the
spectrum than those for $\eta=0.5$, $Re=125$.

Our study is restricted to the azimuthal wavenumber $m=\alpha=0$,
axial wavenumber $\beta=\pi/2$ and angular velocity ratio $\mu=-1$.
Previous results on Couette flows lead us to believe that these values
of $(\alpha,\beta)$ may be representative of a fairly large portion of
parameter space, since both the maximal optimal growth rate in plane
Couette flow and the maximal linear growth rate in Taylor Couette flow
are close to those achieved for $\alpha=0$, $\beta=\pi/2$.  Varying
$\mu$, however, leads to major qualitative changes, as it does in
other aspects of Taylor-Couette flow.  Studies encompassing 
a wide range of parameter values \cite{Meseguer} are clearly desirable.

In the words of Faisst \& Eckhardt \cite{Eckhardt},
Taylor-Couette flow provides an "embedding" of plane Couette flow.
Numerical simulations and laboratory experiments have extensively 
documented the way in which plane Couette flow changes as its single
non-dimensional parameter, the Reynolds number, is increased. 
Taylor-Couette flow provides an ensemble of
other parameter paths along which to approach or to step back from 
plane Couette flow. Our hope is that this preliminary study
and that of \cite{Meseguer} 
of transient growth and pseudospectra in Taylor-Couette flow, 
will help to increase understanding of both Taylor-Couette flow
and of the effects of non-normality.

{\Large \bf Acknowledgments}

We gratefully acknowledge Thomas Wright for use of his code for
calculating pseudospectra, and Satish Reddy for use of his code for
calculating optimal growth.  PJS thanks Patrick Huerre and the people
at LadHyX for their warm hospitality during his sabbatical stay.

\end{document}